\documentclass[aps,pre,twocolumn,showpacs,superscriptaddress]{revtex4}
\usepackage{amsmath}
\usepackage{amsfonts}
\usepackage{amssymb}
\usepackage{epsfig}
\usepackage{graphicx}
\renewcommand{\phi}{\varphi}

\newcommand{\pdif}[2]{\frac{\partial #1}{\partial #2}}
\newcommand{\be}{\begin{equation}}
\newcommand{\ee}{\end{equation}}
\newcommand{\ba}{\begin{align}}
\newcommand{\ea}{\end{align}}
\newcommand{\ave}[1]{\langle {#1} \rangle}
\newcommand{\phij}{\varphi_J}

\begin{document}

\title{Thermal fluctuations, mechanical response, and 
hyperuniformity in jammed solids}
   
\author{Atsushi Ikeda}
\affiliation{Fukui Institute for Fundamental Chemistry, 
Kyoto University, Kyoto, Japan}

\author{Ludovic Berthier}
\affiliation{Laboratoire Charles Coulomb, 
UMR 5221 CNRS-Universit\'e de Montpellier, Montpellier, France}

\date{\today}

\begin{abstract}
Jamming is a geometric phase transition occurring in 
dense particle systems in the absence of temperature. 
We use computer simulations to analyse the effect of thermal 
fluctuations on several signatures of the transition.
We show that scaling laws for bulk and shear moduli only become 
relevant when thermal fluctuations are extremely small, and propose their 
relative ratio as a quantitative signature of jamming criticality.
Despite the nonequilibrium nature of the transition, we find that thermally 
induced fluctuations and mechanical responses obey 
equilibrium fluctuation-dissipation
relations near jamming, provided the appropriate fluctuating 
component of the particle displacements is analysed. 
This shows that mechanical moduli can be directly measured from particle 
positions in mechanically unperturbed packings, and suggests
that the definition of 
a ``nonequilibrium index'' is unnecessary for amorphous materials.
We find that fluctuations of particle displacements are spatially 
correlated, and define a transverse and a longitudinal correlation
lengthscales which both diverge as the jamming transition is approached. 
We analyse the frozen component of density fluctuations 
and find that it displays signatures of nearly-hyperuniform behaviour 
at large lengthscales. This demonstrates 
that hyperuniformity in jammed packings 
is unrelated to a vanishing compressibility and explains why 
it appears remarkably robust against temperature and density variations.
Differently from jamming criticality, obstacles preventing the
observation of hyperuniformity in colloidal systems do not originate 
from thermal fluctuations.
\end{abstract}

\pacs{05.10.-a, 05.20.Jj, 64.70.qj}

\maketitle

\section{Introduction}

A jamming transition \cite{chapter,martin,PZ} 
occurs when it becomes too difficult to compress any further a 
dense assembly of hard objects, whose compressibility then vanishes. 
Remarkably, the same transition also 
controls the loss of mechanical rigidity observed when soft 
particles are decompressed in the absence of thermal fluctuations,
such as foams or emulsion droplets, whose bulk modulus then 
vanishes. In both cases, the key variable controlling the 
physics is the particle connectivity, 
the jamming transition corresponding to the isostatic situation where 
just enough contacts are present to 
insure mechanical stability \cite{mouzarkel,tom,roux,epitome}. 
The distance to isostaticity
controls the divergence of mechanical moduli observed when 
compressing hard particles and their vanishing when decompressing soft 
particles \cite{epitome,wyart}. The deep connection between geometry and 
mechanical responses shows that the criticality observed in 
the vicinity of the transition follows from the unambiguous 
identification of particle contacts.
Therefore, when thermal fluctuations are present, for instance when
considering colloidal particles (such as microgels, emulsions, PMMA 
colloids), the contacts can be `blurred' 
by thermal agitation and cannot be resolved, which challenges the
possibility to observe jamming criticality in experiments.  
In other words, {\it a thermal system cannot know how far it is 
from isostaticity}, and the associated criticality is 
easily destroyed by temperature \cite{ikeda2}.  
  
In previous work, the role of thermal fluctuations near jamming
has been explored to understand the influence of finite 
temperatures on various physical quantities such 
as microscopic dynamics, microstructure, contact number, mechanical 
properties~\cite{brito,brito2,ikeda2,hugo1,hugo2,ikeda1,ikeda3,wyart2014,wyart2015,vestige,xu1,bertrand}. 
In particular, a computer study of the single particle dynamics 
revealed the existence of a very narrow region in the 
(density, temperature) phase diagram where jamming criticality 
can be observed, which excludes most colloidal studies to 
date~\cite{ikeda2} . 
More recent experiments have concentrated 
on collective static properties, such as mechanical shear and bulk moduli and 
structure factors, and the results were analysed using 
power laws that are valid, strictly speaking, for fully athermal 
systems \cite{yodh1}. To assess  the validity of this description, one needs to
extend the analysis of Ref.~\cite{ikeda2} 
to mechanical moduli to understand whether 
their critical behaviour is robust against thermal fluctuations. 
The first goal of our work is to analyse the effect 
of thermal fluctuations on mechanical moduli near jamming.  

Another property characterizing jammed packings is their 
hyperuniformity, which was revealed by analysing the large-distance
scaling of volume fraction fluctuations \cite{donev2}. For monodisperse 
spherical particles of diameter $\sigma$, this reduces to studying the 
ordinary static structure factor, $S(k)$, whose low-wavevector 
behavior obeys a nontrivial, characteristic 
linear behaviour, $S(k \sigma \ll 1) 
\sim k$, which shows that density fluctuations are suppressed
at large scale~\cite{donev2,torquato2003}. 
This behavior has been observed numerically 
in particle packings prepared {\it exactly at the jamming 
transition} \cite{donev2,torquato2003,zac,berthier,silbert3},
and experimentally in athermal granular 
materials \cite{berthier}. Experiments 
performed with colloidal particles appear challenging and report only 
very weak signs of hyperuniformity \cite{weeks1,kurita,dreyfus}. 
A possible explanation
could be that hyperuniform behaviour is blurred by thermal
fluctuations, as are other signatures of the jamming transition.
However, hyperuniformity is a property of the packings at large 
lengthscale and the above argument regarding the resolution of 
particle contacts is not obviously relevant. Therefore, if hyperuniformity 
were affected by thermal fluctuations acting 
at the (vanishingly small) contact lengthscale, it would 
directly establish that hyperuniformity is another critical 
property associated to the jamming transition. 
The second goal of our work is to test whether hyperuniformity 
is robust against thermal fluctuations, and, more fundamentally, 
whether hyperuniformity is deeply related to the jamming
transition, or is instead a distinct phenomenon.   

Thermal fluctuations in jammed packings not only raise 
practical issues about experimental observations, they also
pose fundamental challenges related to the nonequilibrium 
nature of the jamming transition. As mentioned above, mechanical moduli 
display power law behaviour near jamming at zero temperature. However,
for materials at thermal equilibrium, mechanical response functions are 
directly related to mechanical fluctuations induced 
by thermal motion and thus to equilibrium structure 
factors through fluctuation-dissipation 
relations \cite{chaikin-lubensky,hansen}. 
Near the nonequilibrium jamming transition at finite 
temperatures, two behaviors are then possible:

{\it 1) Fluctuations and responses do not obey equilibrium relations}, 
so that mechanical moduli and structure factors have independent 
density and temperature dependences. This hypothesis 
suggests that it could be useful to introduce novel variables 
to quantify deviations from equilibrium relations,
such as nonequilibrium index \cite{torquato2,torquato3} 
or effective temperatures \cite{leticia,jorge},  
generically defined as ratios between fluctuations and responses.
In that case, structure factors live an independent life from 
mechanical responses, and they may display an independent set of critical
properties, but they may also have unremarkable behaviour near jamming. 
This general hypothesis has been advocated in particular in 
Refs.~\cite{torquato2,torquato3}, where a diverging nonequilibrium index and a 
diverging nonequilibrium lengthscale were defined from fluctuations 
and responses of hard sphere systems approaching jamming \cite{torquato2}, 
and later extended to generic amorphous solids \cite{torquato3}.  

{\it 2) Fluctuations and responses obey equilibrium 
relations,} and the critical behaviour of mechanical 
moduli should have a counterpart in fluctuations 
of particle positions and structure factors. In that case, 
if thermal fluctuations are finite (but still sufficiently small 
that they do not blur the jamming criticality!), 
interesting critical behaviour should be observed in density fluctuations
of mechanically unperturbed packings. In particular, one may expect the 
emergence of diverging lengthscales in collective structure factors of jammed 
materials. 

The third goal of our work is to decide which of the two above 
scenarios is valid, and whether interesting lengthscales 
and nonequilibrium indicators
emerge from the analysis of structure factors.  

To achieve our three main goals, we use computer simulations of a simple 
model of soft harmonic particles \cite{durian}
to analyse the influence of thermal fluctuations on mechanical 
responses and structure factors in the vicinity of the jamming transition. 
Harmonic spheres are convenient 
because they allow studies of multiple, 
experimentally relevant, routes to jamming in the 
(density, temperature) phase diagram 
\cite{epitome,ikeda1,durian,teitel,tom1,tom2}. Therefore, a single model
provides us with decisive answers to the three sets of questions 
mentioned above, that can be summarized as follows.

(i) We find that mechanical moduli are as sensitive to thermal fluctuations  
as single particle dynamics and their associated power law behaviour 
is not a good starting point to theoretically describe existing
colloidal experiments.

(ii) By contrast, hyperuniformity is extremely robust to 
the addition of thermal perturbations, and even to changes
in packing fraction, suggesting that it should in fact be far easier 
to observe in experiments than the jamming criticality, even though 
the present state of the literature suggests the opposite. We also conclude
that hyperuniformity bears no deep relation
to the jamming transition, and in particular we show that 
it is fully unrelated to the critical 
behavior of the mechanical compressibility. 

(iii) Equilibrium fluctuation-dissipation relations are
perfectly obeyed near jamming, suggesting it is unnecessary 
to define quantitative indicators for the degree of 
`nonequilibriumness' near jamming. It also implies that 
structure factors display critical properties and 
reveal diverging lengthscales, that we define, analyse, 
and compare to previously studied critical lengthscales. 

This article is organised as follows. 
In Sec.~\ref{model} we define the model and our numerical strategy. 
In Sec.~\ref{mechanical}, we analyse the behaviour of mechanical
moduli.
In Sec.~\ref{static} we define and analyse the behavior of structure 
factors and their associated lengthscales. 
In Sec.~\ref{hyperuniformity} we discuss the hyperuniformity 
of jammed packings. 
In Sec.~\ref{summary} we summarize and discuss our results. 

\section{Model and simulation}
\label{model}

We consider a system of monodisperse harmonic spheres, 
interacting through a pairwise potential \cite{durian}, 
\be
v(r_{ij}) =  \frac{\epsilon}{2} 
(1-r_{ij}/\sigma)^2  \Theta(\sigma - r_{ij}) ,
\ee 
where $\Theta(x)$ is the Heaviside function, $r_{ij}$ is the distance 
between particles $i$ and $j$, and $\sigma$ 
is the particle diameter. 
Throughout this work, length, energy, temperature and mechanical moduli 
are measured in units of $\sigma$, $\epsilon$, $\epsilon/k_B$, and 
$\epsilon/\sigma^3$, respectively. 

We use molecular dynamics simulations \cite{allen-tildesley}
to compute the mechanical moduli 
and static structure factors of the system at finite temperature. 
The setting of the calculations is essentially similar to our previous 
work~\cite{ikeda2}. 
We first generate a random configuration of $N=64,000$ particles in a 
simulation box with periodic boundary conditions. 
The linear dimension of the box $L$ is adjusted to realize the packing 
fraction fraction $\varphi=0.80$, where $\phi = \pi \sigma^3 N / (6L^3)$. 
Starting from this configuration, we perform 
molecular dynamics simulations at $T=10^{-5}$, 
where we integrate Newton's equations of motion using velocity 
rescaling to control the temperature. 
This can be seen as an extensive aging of the system starting 
from $T = \infty$ down to $T=10^{-5}$. We find that temperature is low
enough that aging dynamics eventually stops
and the energy and average particle positions reach well-defined values
that do not depend on waiting time any longer. 
After this long annealing of the system, we change 
the density and temperature to the desired values smoothly, 
letting the system relax at each state point before taking any measurement. 
This protocol allows us to study essentially the same particle packing
at different densities and temperatures. We do not study temperatures
larger than $T=10^{-5}$, and in the studied regime particle diffusion 
and rearrangements can be safely neglected.  
 
At each state point, we then perform molecular dynamics simulations 
to calculate the mechanical moduli and static structure factors, 
where we integrate Newton's equations of motion in the $NVE$ ensemble, 
i.e. without thermostat.  
We denote the long-time average in these calculations with brackets, 
$\ave{\cdots}$.  
For the particular configuration which is analysed extensively in this work, 
the jamming density is $\phij \simeq 0.648$. 

For the specific purposes of Sec.~\ref{hyperuniformity}, we also 
calculate the static 
structure factor of jammed harmonic spheres with 
a larger number of particles, $N=512,000$, at strictly zero temperature. 
To this end, we generate a random configuration of particles in 
a simulation box with $\varphi=0.80$, and 
then apply the FIRE algorithm to minimize the potential energy of 
the system at this density~\cite{fire}.  
Starting from this jammed configuration, we decrease the density 
by small steps and minimize the potential energy after each step 
to obtain a series of jammed particle configurations over a range
of densities \cite{epitome}. To increase the statistics of the results obtained 
for these zero-temperature packings, we followed this procedure 
starting from 8 independent random configurations, and finally
averaged the results over these independent runs. For this 
series of simulations, we find that $\phi_J \simeq 0.64571 \pm 0.00012$,
where the errorbar indicates the standard deviation among independent
packings. Averaging over those configurations is therefore 
accurate as long as the distance to the jamming density is larger 
than $|\phi - \phi_J| \approx 0.0001$.

\section{Mechanical moduli and jamming criticality}
\label{mechanical}

In this section we analyse the temperature and density dependences 
of bulk and shear moduli of harmonic spheres in the vicinity of the 
jamming transition occurring at $(T=0, \phi=\phi_J)$. 

\subsection{Bulk modulus}

\begin{figure}
\psfig{file=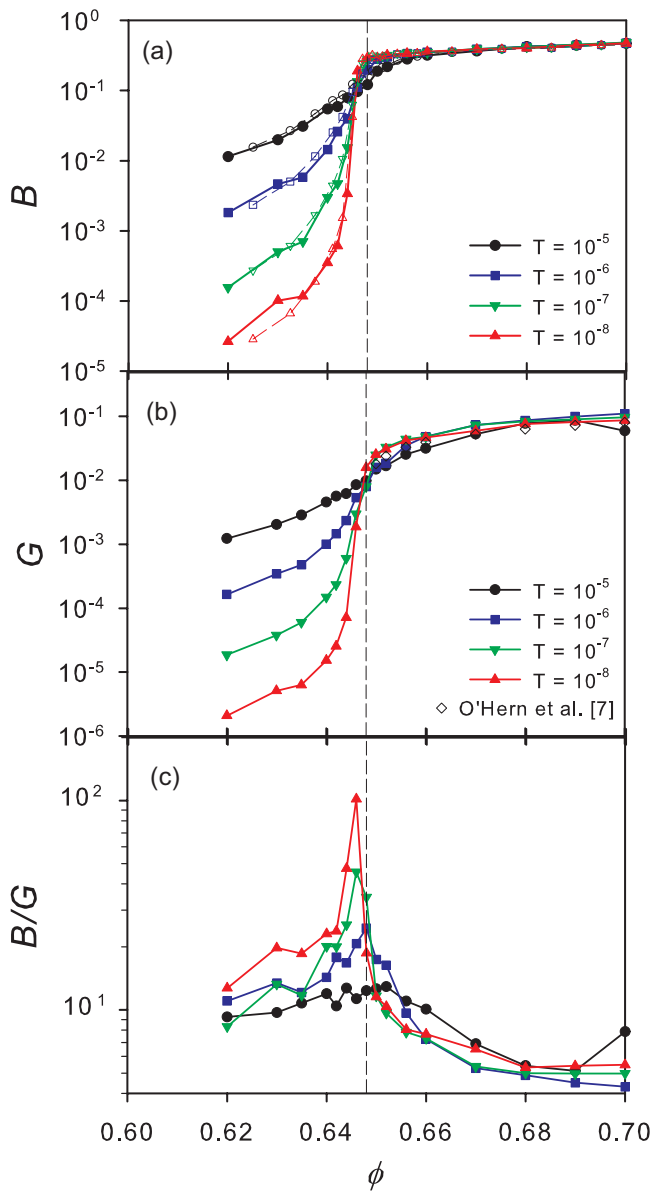,width=8.5cm,clip}
\caption{(Color online) Density dependence of various quantities
for different temperatures indicated in the label.
(a) Bulk modulus.
Results from the pressure derivative,
Eq.~(\ref{dpdv}), are shown with open symbols and dashed lines,  
results from the fluctuation formula, Eq.~(\ref{p2}), are shown  
with filled symbols and solid lines.
(b) Shear modulus obtained from Eq.~(\ref{Gdef}). 
Open diamonds indicate data for $T=0$ 
obtained from response functions~\cite{epitome}.
(c) Ratio of the bulk to shear modulus.
The vertical dashed line indicates the location of the $T=0$
jamming transition at $\phi_J \simeq 0.648$. 
\label{bulk}}
\end{figure}

We start our analysis with  the calculation of the bulk modulus. 
The isothermal bulk modulus $B$ quantifies the resistance of the 
system to compression. Its definition then naturally
involves the pressure (noted $P$) derivative of the 
volume (noted $V$),
\begin{eqnarray}
B = -V \Bigl( \pdif{P}{V} \Bigr)_T. \label{dpdv}
\end{eqnarray}
We first calculate $B$ through the response formula Eq.~(\ref{dpdv}), 
where the pressure $P$ is calculated from the virial formula 
\be
P = \rho T + \frac{\ave{W}}{V},
\ee 
where $W = \sum_{ij} r_{ij} v'(r_{ij})/3$ 
is the virial \cite{hansen}. The bulk modulus is of course inversely
proportional to the isothermal compressibility, 
$B = 1 / \chi_T$. In practice we measure the pressure for various 
densities, and estimate the derivative in Eq.~(\ref{dpdv})
using finite differences, which suggests that compressions or 
decompressions yield the same results.  
The numerical results are shown with open symbols and dashed lines 
in Fig.~\ref{bulk}. 

The density dependence of the bulk modulus strongly depends on the temperature. 
At lower temperature, e.g. $T=10^{-8}$, the bulk modulus increases 
very sharply with density when $\varphi < \phij$, 
and becomes essentially density-independent when $\varphi > \phij$. 
This behavior can be understood as a smooth crossover between the jamming of 
Brownian hard spheres 
and the unjamming of non-Brownian soft spheres.
For $\varphi < \phij$ and $T \to 0$, 
the particles have vanishing overlaps 
and essentially explore hard sphere configurations. 
The pressure of hard spheres diverges at the jamming 
transition as $P \sim T (\phij - \varphi)^{-1}$, and as a result 
the bulk modulus behaves as~\cite{donev1} 
\be 
B \sim T (\phij - \varphi)^{-2} ,
\label{Bdiv}
\ee
which underlies both the critical nature of the transition 
and the entropic origin of solidity in hard particle systems. 
Our numerical results for $\varphi < \phij$ can be well fitted 
with this power-law divergence. Equivalently, 
Eq.~(\ref{Bdiv}) implies that the isothermal compressibility 
$\chi_T$ vanishes quadratically with $(\phi_J-\phi)$ in this regime.
 
On the other hand, for $\varphi > \phij$ at lower temperatures, 
particles may have finite overlaps, and thermal fluctuations play 
a small role. Thus, the system corresponds to non-Brownian 
soft spheres. For this system, the pressure emerges continuously 
at the jamming transition, $P \sim (\varphi - \phij)$, 
and thus the bulk modulus is expected to be~\cite{epitome} 
\be
B \sim \mbox{const}.
\label{Bright}
\ee 
Again, this behavior is in good agreement 
with the results shown in Fig.~\ref{bulk}. 

When temperature is increased above $T=10^{-8}$, 
the difference of behaviour observed on both sides 
of the transitions becomes smaller, and for the largest 
studied temperature, $T=10^{-5}$, the bulk modulus is a smooth function 
of density across $\phi_J$. This shows that systems characterized
by $T=10^{-5}$ (in units of the particle softness $\epsilon$ \cite{ikeda3}) 
are unable to reveal signs of the underlying jamming transition
at $T=0$. In particular, the solid behaviour of these  
systems is better interpreted as resulting from 
hitting a glass transition line $T_g(\phi)$ \cite{rmp}.  
This qualitative behavior is consistent with our previous 
discussion of the single particle dynamics~\cite{ikeda2}. 

We now use a different approach to compute the bulk modulus
which does not involve a response function, but stems instead 
from studying thermal fluctuations in a mechanically unperturbed 
material. Within the framework of equilibrium statistical mechanics, 
the bulk modulus can be directly related to the fluctuations of the pressure. 
In the $NVE$ ensemble that we use to compute the pressure fluctuations, 
the formula for the isothermal bulk modulus reads~\cite{allen-tildesley} 
\begin{eqnarray} 
B = P + \frac{\ave{W_2}}{V} - \frac{\ave{P^2} - \ave{P}^2}{T}V + \frac{2}{3} \rho T - \frac{T \gamma_V^2}{\rho c_V}, 
\label{p2}
\end{eqnarray}
where $W_2 = \sum_{ij} (r_{ij} v'(r_{ij}) + r_{ij}^2 v''(r_{ij}) )/9$ 
is the hyper-virial, $c_V$ is the specific heat per particle, 
and $\gamma_V$ is the thermal pressure coefficient~\cite{allen-tildesley}. 
The last term in Eq.~(\ref{p2}) arises because we work 
in the microcanonical ensemble where the energy is conserved.  

We have calculated the bulk modulus through Eq.~(\ref{p2}), and
report the results as filled symbols in Fig.~\ref{bulk}.
Clearly, the results are in excellent agreement 
with the ones obtained from the response formula, Eq.~(\ref{dpdv}). 
Therefore, we conclude that {\it fluctuations and response
functions yield identical results} for repulsive colloidal 
particles near jamming over a broad range of densities 
and temperatures. The natural interpretation is that  
equilibrium relations appear satisfied because deeply jammed solids
dynamically explore a restricted portion of the configurational space located
near a metastable amorphous configuration. In other words, 
the system is locally in equilibrium, even though
ergodicity is globally broken. Using the language of 
the two-temperature scenario for aging glasses \cite{leticia}, the thermal 
fluctuations that are probed in the present system correspond
to the fast degrees of freedom that appear locally 
equilibrated at the temperature of the thermal bath.
This physical perspective justifies why it is unnecessary to introduce 
an effective temperature for the slow degrees of freedom, because
these are completely frozen in the type of analysis that we perform.
In other words, vibrational motion does not reveal the nonequilibrium
nature of the glass. 

Our results seem to contradict previous work \cite{torquato2}
introducing a nonequilibrium index $X$ to quantify deviation
from equilibrium behaviour between the bulk modulus $B$
(measured as a reponse function) and the small-wavevector limit
of the static structure factor, $S(k \to 0)$, which 
reduces to the isothermal compressibility $\rho T \chi_T$ for 
equilibrium fluids. In Sec.~\ref{static}, we clarify the 
relation between structure factors, pressure fluctuations, 
and compressibility, and show that the physical content of 
the ``nonequilibrium'' index introduced in Ref. \cite{torquato2}
can in fact be fully understood in terms of equilibrium 
fluctuation-dissipation relations.  

\subsection{Shear modulus}

We now turn to the analysis of the shear modulus, $G$. 
There are several ways to compute the shear modulus in numerical
simulations~\cite{allen-tildesley}. 
The first option is to use global fluctuations. Just as  
the bulk modulus can be determined from the fluctuations 
of the pressure, the shear modulus can be obtained from 
the fluctuations of the shear stress. 
We have first tried to use this approach to calculate the shear modulus,
but found that an accurate determination of $G$ is not easy 
because the fluctuation formula requires to take the difference 
between large numbers which largely cancel and have important 
statistical fluctuations. 

To overcome this problem, we use the alternative
method introduced in Ref.~\cite{klix}. 
In this approach, the shear modulus is calculated as the $k \to 0$ limit of the 
the correlation function of the transverse displacement $S_T(k)$,
\begin{eqnarray}
G = \lim_{k \to 0} \frac{\rho T}{S_T(k)}.
\label{Gdef}
\end{eqnarray}
The precise definition and detailed analysis of $S_T(k)$ will be given 
in Sec.~\ref{static}, see Eq.~(\ref{defST}). For the moment,
we simply notice that the determination of $G$ from 
Eq.~(\ref{Gdef}) clearly stems from spontaneous 
fluctuations, and this approach thus 
differs from earlier determinations based on response functions~\cite{epitome}. 
Here, we concentrate on the temperature and density dependences of 
the shear modulus $G$, and report our results in Fig.~\ref{bulk}.
Note that this definition of the shear modulus does not 
require testing the validity of linear response, and does not  
depend either on the chosen direction for shearing.
Although the shear modulus measured as a response 
function may depend on the direction of shear~\cite{dagois},
all the directions of the shear modes are averaged out in the 
definition of $S_T(k)$ that we use in Eq.~(\ref{defST}).

First, we check the validity of the fluctuation formula Eq.~(\ref{Gdef}). 
To this end we compare our results to the shear modulus obtained from 
the response function in Fig.~\ref{bulk}. 
Although the available data is limited to the density above 
$\phij$ at $T=0$~\cite{epitome}, 
our results at lower temperature are quantitatively the same as 
the data from the response function at $T=0$. 
This confirms that fluctuations and response functions 
yield identical results for the shear modulus as well. A similar
agreement between response and correlations for the shear modulus
was reported in other glassy systems~\cite{klix,baschnagel},
which appears as a robust result.

The overall behavior of the shear modulus is qualitatively similar to the 
one of the bulk modulus. At lower temperature, the shear modulus also 
increases very sharply with $\phi$ below the jamming density, 
and has a more modest density dependence above jamming. 
A closer look to the numerical data indicates that the density 
dependence of $G$ is more pronounced above jamming than the one 
of $B$. Similarly to $B$, the sharp features of the 
shear modulus disappear rapidly when temperature is increased
above $T=10^{-8}$, and again the density dependence is very smooth
when $T \gtrsim 10^{-6}$. This indicates that the characteristic 
critical laws associated to the shear modulus near the 
jamming transition are easily smeared out by thermal fluctuations
as well. 

The low-temperature crossover behaviour observed for $G$ is again
the signature of the zero-temperature criticality associated to the jamming 
transition. For $\phi < \phi_J$ and $T \to 0$, the system 
explores the divergence of the shear modulus of Brownian 
hard spheres approaching jamming, which follows 
\be
G \sim T (\phi_J - \phi)^{-\kappa},
\label{Gleft}
\ee 
where $\kappa \approx 1.41$ is a non-trivial critical 
exponent~\cite{yoshino,naturecomm,wyart2014}. 
On the other side of the jamming transition, 
jammed harmonic spheres lose shear rigidity as the jamming 
density is approached from above~\cite{wyart2015}, 
\be
G \sim (\phi - \phi_J)^{1/2}. 
\label{Gright}
\ee  
Although our numerical results are consistent 
with Eqs.~(\ref{Gleft}, \ref{Gright}), they are not precise enough 
to confirm that the exponent $\kappa$ is different from 
a previous estimate $\kappa = 3/2$~\cite{brito}, which is only marginally 
different from its recently predicted value $\kappa \approx 1.41$. 

\subsection{Ratio $B/G$ of bulk to shear modulus: a signature 
of jamming criticality}

Whereas we noted that both $B$ and $G$ show qualitatively similar
sharp features in the vicinity of the jamming transition, 
we also stated that the precise values of the exponents 
characterizing their power law behavior are different. 
These differences stem from the fact that  
the behaviour of the bulk modulus $B$ can be understood 
from the evolution of the pressure, whereas the behavior of $G$ is ruled 
by the evolution of the response of the system to a shear deformation. 
It is a specific signature of the jamming transition 
that responses to shear and to compression differ 
maximally for isostatic packings 
\cite{chapter,martin,epitome,matthieu,martin3}.  

Therefore, to clearly detect a quantitative sign of the jamming 
criticality, it is useful to analyse the behavior of the ratio
$B/G$, which becomes infinite at the critical point. 
We combine our finite temperature data for $B$ and $G$ to 
follow the density dependence of $B/G$ for various 
temperatures in Fig.~\ref{bulk}. 
For our lowest temperature, $T=10^{-8}$, we find 
that $B/G$ is of order 5-10 far from the jamming density, 
but has a sharp maximum of order $B/G \approx 100$ when 
$\phi \approx \phi_J$. This behavior should be interpreted as a 
smooth version of the zero-temperature density dependence, which follows 
from Eqs.~(\ref{Bdiv}, \ref{Bright}, \ref{Gleft}, \ref{Gright}):
\begin{eqnarray}
B/G & \sim & (\phi_J - \phi)^{\kappa-2}, \quad \phi < \phi_J, \\
 & \sim &  (\phi - \phi_J)^{-1/2}, \quad \,  \, \phi > \phi_J,
\label{BsurG}
\end{eqnarray}
where $\kappa-2 \approx -0.59$. Notice that the behaviour of $B/G$
is now more symmetric around the jamming transition as the 
temperature prefactor disappears from the ratio $B/G$, 
but the critical exponents slightly differ on both sides 
of the transition (the divergence should be sharper 
for $\phi < \phi_J$). Note also 
that the behavior of $B/G$ is quantitatively analogous 
to the behavior of the adimensional mean-squared displacement
defined in Ref.~\cite{ikeda2}. Therefore, this 
figure demonstrates that the impact of thermal fluctuations 
on mechanical moduli and on single particle dynamics 
is actually identical, and {\it mechanical moduli are in fact equally fragile 
against Brownian motion}. 

We interpret the smoothened version of the 
symmetric divergence described by Eq.~(\ref{BsurG}) observed 
for $T=10^{-8}$ in Fig.~\ref{bulk} as a `thermal vestige'
of the jamming transition \cite{vestige}, which is equivalent to 
the adimensional mean-squared displacement defined in Ref.~\cite{ikeda2}. 
These two quantities 
are both direct signatures of jamming criticality and should therefore 
be contrasted with a non-monotonic behaviour of the pair correlation
functions \cite{vestige,hugo1}
which is instead a more general consequence of the particle 
softness and, as such, survives arbitrarily far 
from the critical point \cite{hugo3}. Therefore
we suggest that the observation of a large $B/G$ ratio is a 
genuine sign that a particular material lies in the critical 
regime of the jamming transition, 
whereas a density-maximum in the pair correlation function is not.
 
When the temperature is increased, the non-monotonic 
density dependence of $B/G$ is rapidly erased by thermal fluctuations.
For $T=10^{-5}$, we observe that $B/G$ is nearly independent of the density
and has a value of order $5-10$ at all $\phi$. 
These findings directly confirm that the jamming criticality is 
rapidly smeared out by thermal fluctuations. We also notice 
that even for very low temperatures, the range of densities where
anomalous behavior associated to jamming can be observed
is extremely narrow. These conclusions, obtained from 
the analysis of mechanical moduli, are in full agreement with previous
conclusions drawn from the analysis of the mean-squared 
displacements \cite{ikeda2}.

In Ref.~\cite{yodh1}, the bulk and shear moduli 
of a two-dimensional assembly of soft microgels 
were analysed, and their density evolution 
interpreted in terms of the power laws associated 
to the zero-temperature jamming criticality. Previous analysis of 
similar microgel systems has shown that
these particles are quite soft, so that thermal
fluctuations are of order $T \approx 10^{-6} - 10^{-4}$, depending
on the precise experimental system \cite{ikeda3,yodh2}. The data reported 
in Ref.~\cite{yodh1} show that the ratio $B/G$
is $B/G \approx 3$ with a very weak density dependence. This is 
very much consistent with the physical interpretation 
that this system is far from being critical, which reinforces 
the general conclusion that very soft microgel systems are not 
good experimental systems 
to reveal thermal vestiges of the jamming transition.
In Ref.~\cite{ikeda3}, we suggested that emulsion droplets
might be better suited for this task, as recently confirmed experimentally
\cite{emulsion}.
 
\section{Static structure factors and 
diverging lengthscales}
\label{static}

In this section we define and study a number of 
static structure factors that can be probed in 
kinetically arrested colloidal materials in the 
presence of thermal fluctuations. From their low-wavevector 
analysis, we define lengthscales that diverge as the $T=0$ jamming 
transition is approached in the $(T, \phi)$ plane.  

\subsection{Definitions of structure factors}
\label{definitions}

Because we know the position of each particle 
in each configuration, we can define a number of static structure 
factors from our particle packings at finite temperatures. 

The standard definition of the static structure factor is given by   
\begin{eqnarray}
S(k) = \frac{1}{N} \ave{ \rho_{\vec{k}}  \rho_{-{\vec k}} }, 
\label{defS}
\end{eqnarray}
where we have defined the Fourier transform of the density field as 
\be \rho_{\vec{k}} = \sum_j \exp (-i \vec{k} \cdot \vec{R}_j),
\ee 
with $\vec{R}_j$ the position of particle $j$.
We assume that all our packings are isotropic so that 
structure factors only depend upon wavevectors through
their moduli $k = | \vec{k} |$. 

For a simple fluid at thermal equilibrium, the zero wavevector 
limit of $S(k)$ is directly related to the bulk modulus, and we 
have 
\begin{eqnarray}
\lim_{k \to 0} S(k) =  \frac{\rho T}{B} \ \ \ \mbox{(Fluid).} \label{liqb}
\end{eqnarray}
This relation is only correct for 
liquid states~\cite{hansen}. We shall see below that 
a different analysis is needed for jammed solids. 

We consider systems that are lacking crystalline order but 
are nevertheless kinetically arrested. This  implies that 
translational invariance is actually broken, and that we consider 
instead materials with long-range ``amorphous order'' to use 
the language of glass theories \cite{rmp,szamel}. In practice, this means that 
the particle positions, and therefore, density 
fluctuations, can be naturally decomposed into 
two different contributions, from which two 
distinct structure factors can be defined: 
\begin{eqnarray}
S_{\delta}(k) &=& \frac{1}{N} \ave{\delta \rho_{\vec{k}} 
\delta \rho_{-{\vec{k}}} }, \label{defSdelta} \\  
S_0(k) &  = & \frac{1}{N} \ave{\rho_{\vec{k}}} 
\ave{\rho_{-{\vec{k}}}},  \label{defS0} 
\end{eqnarray}
with $\delta \rho_{\vec k} = \rho_{\vec k} - \ave{\rho_{\vec k}}$. 
From the definitions (\ref{defS}, \ref{defSdelta}, \ref{defS0})
it is obvious that  
\begin{eqnarray}
S(k) = S_{\delta}(k) + S_0(k),
\label{sum}
\end{eqnarray}
showing that we have obtained a decomposition 
of the structure factor in terms of a fluctuating part, 
$S_{\delta}(k)$, and a configurational part, $S_0(k)$. 
Physically, $S_0(k)$ represents the structure factor associated 
to the averaged position of the particles and is essentially
independent of temperature and weakly dependent of density; 
it is an `inherent' property of the amorphous packing \cite{SW}.  
By contrast, $S_{\delta}(k)$ represents the structure factor 
associated to the fluctuations of the particles away from 
their average positions, and a strong 
temperature dependence is expected for this contribution, which 
should for instance vanish as $T\to 0$ for $\phi>\phi_J$, 
when particles stop moving completely.  

When translational invariance is broken, as in crystals and glasses, 
the bulk modulus of the system is no longer related to $S(k)$ 
as in Eq.~(\ref{liqb}), but to the fluctuation part 
$S_{\delta}(k)$~\cite{chaikin-lubensky}. 
Within a conventional conventional elasticity theory 
where the elastic moduli are assumed to be independent 
of wavevector, 
an elastic body is characterized by longitudinal plane waves with the 
dispersion relation $\omega = k \sqrt{(B+\frac{4}{3}G)/\rho}$. 
These plane waves are thermally excited and follow the equipartition 
law, so that  
the fluctuating part of the density fluctuations is given by~\footnote{To 
arrive to this expression, 
we first evaluate $S_L(k)$ using the dispersion relation, 
and used Eq.~(\ref{sim}) to estimate $S_{\delta}(k)$. 
Details of the evaluation of $S_L(k)$ can be found in 
pp.~321-322 of Ref.~\cite{chaikin-lubensky}.} 
\begin{eqnarray}
S_{\delta}(k) = \frac{\rho T}{B+\frac{4}{3}G} \ \ \ 
\mbox{(Continuum solid).} \label{fdtsk}
\end{eqnarray}
This is the fluctuation formula appropriate for connecting 
a static structure factor to the bulk modulus in a solid state.
It is obviously distinct from the formula valid for fluids, 
Eq.~(\ref{fdtsk}). Notice that the distinction between the two
formula stems from the fact that translational invariance 
is broken (in solids) or not (in fluids), but both formula 
rely on the fact that the system (fluid or solid) obeys the 
rules of equilibrium thermodynamics. 

It is also useful to provide a dynamic interpretation of the
decomposition in Eq.~(\ref{sum}). Because the dynamics is arrested 
and the system only probes thermal fluctuations near a
given metastable state, the dynamic structure factor 
does not decay to zero at long times. The intermediate scattering 
function is $F(k,t) = \ave{ \rho_{\vec{k}}(0)  \rho_{-{\vec k}} (t)} / N$,
so that $F(k,t=0) = S(k)$. In the long-time 
limit, density fluctuations are uncorrelated, and we 
directly find that $F(k,t \to \infty) = S_0(k)$, which
is nothing but the collective Debye-Waller factor.   
Therefore, the fluctuation part of the static structure
can be written as $S_{\delta}(k) = S(k) - F(k, t \to \infty)$,
which quantifies the relaxing part of the density fluctuations. 

In addition, we define two more structure factors associated 
to the particle positions, which rely on the vectorial 
character of the displacement field. 
In solid states, each particle vibrates around its average position. 
We can then define the displacement of each particle as 
$\vec{u}_i = \vec{R}_i - \ave{\vec{R}_i}$, 
and the associated displacement field, expressed 
in the Fourier domain as  
\begin{eqnarray}
\vec{u}_{\vec{k}} = \sum_j 
\vec{u}_j \exp (-i \vec{k} \cdot \ave{\vec{R}_j}). \label{disp} 
\end{eqnarray}
In the Fourier space, we can then decompose the displacement field 
into its longitudinal and transverse parts:
\be 
\vec{u}_{\vec{k}} = \hat{k} u_{L,\vec{k}} + \vec{u}_{T,\vec{k}},
\ee 
where $u_{L,\vec{k}} = \hat{k} \cdot 
\vec{u}_{\vec{k}}$, and $\hat{k} = \vec{k}/|\vec{k}|$ is the unit vector 
in the direction of  $\vec{k}$. 
Using these fields, we can finally define the 
longitudinal and transverse correlation functions
\begin{eqnarray}
S_L(k) & = & \frac{k^2}{N} \ave{u_{L,\vec{k}} u_{L,-\vec{k}}}, \nonumber \\ 
S_T(k) & = & \frac{k^2}{N} \ave{\vec{u}_{T,\vec{k}} \cdot \vec{u}_{T,-\vec{k}}}. 
\label{defST}
\end{eqnarray}
Following again the approach of conventional elasticity theory, 
we now have the following expressions~\cite{chaikin-lubensky}
\begin{eqnarray}
S_L(k) & = & \frac{\rho T}{B+\frac{4}{3}G}, \\ 
S_T(k) & = & \frac{\rho T}{G},  \quad \quad \quad 
\ \ \ \mbox{(Continuum solid).} 
\nonumber \label{fdtst} 
\end{eqnarray}
These expressions directly show that we can calculate 
the shear modulus $G$ from the low-wavector limit of 
$S_T(k)$~\cite{klix}. It is the approach that has been employed to obtain the 
data shown in Fig.~\ref{bulk} in Sec.~\ref{mechanical}. 

Before using these definitions we wish to comment that 
the various structure factors defined in this section 
correspond to different mathematical ways 
of decomposing the particle positions into an average 
and a fluctuating contributions. In the first approach, we
performed the decomposition in the Fourier domain, whereas 
the second approach deals with position space. It should therefore 
come as no surprise that the longitudinal part $S_L(k)$ can be 
related to the fluctuating part of the static structure factor 
$S_{\delta}(k)$. In terms of the particle displacements $\vec{u}_i$, 
the fluctuating part of the density field 
can be expressed as
\begin{eqnarray}
\delta \rho_{\vec{k}} = \sum_j (- i \vec{k} \cdot \vec{u}_j 
+ \mathcal{O}(k^2) ) \exp (-i \vec{k} \cdot \ave{\vec{R}_j}).
\end{eqnarray}
Comparing this expression to Eq.~(\ref{disp}), we get $\delta \rho_{\vec{k}} 
\approx -ik u_{L,\vec{k}}$ at the lowest order in $k$. 
Therefore we conclude that at this order
\begin{eqnarray}
S_L(k) \approx S_{\delta}(k), \label{sim}
\end{eqnarray}
and both approaches actually carry equivalent physical content, 
as they should. By a similar reasoning, we find that 
\be
\ave{ \rho_{\vec k} } = \sum_j \exp( -i {\vec k} \cdot 
\ave{{\vec R}_j} ) (1 - k^2 \ave{(\hat{k} \cdot \vec{u}_j)^2}/2 + \cdots ),
\ee
which shows that an accurate estimate of the configurational part 
$S_0(k)$ can be obtained 
by computing the structure factor of the average
particle positions, a method that could prove convenient
for experiments using particle imaging.  

\subsection{Longitudinal fluctuations}
\label{longitudinal}

\begin{figure}
\psfig{file=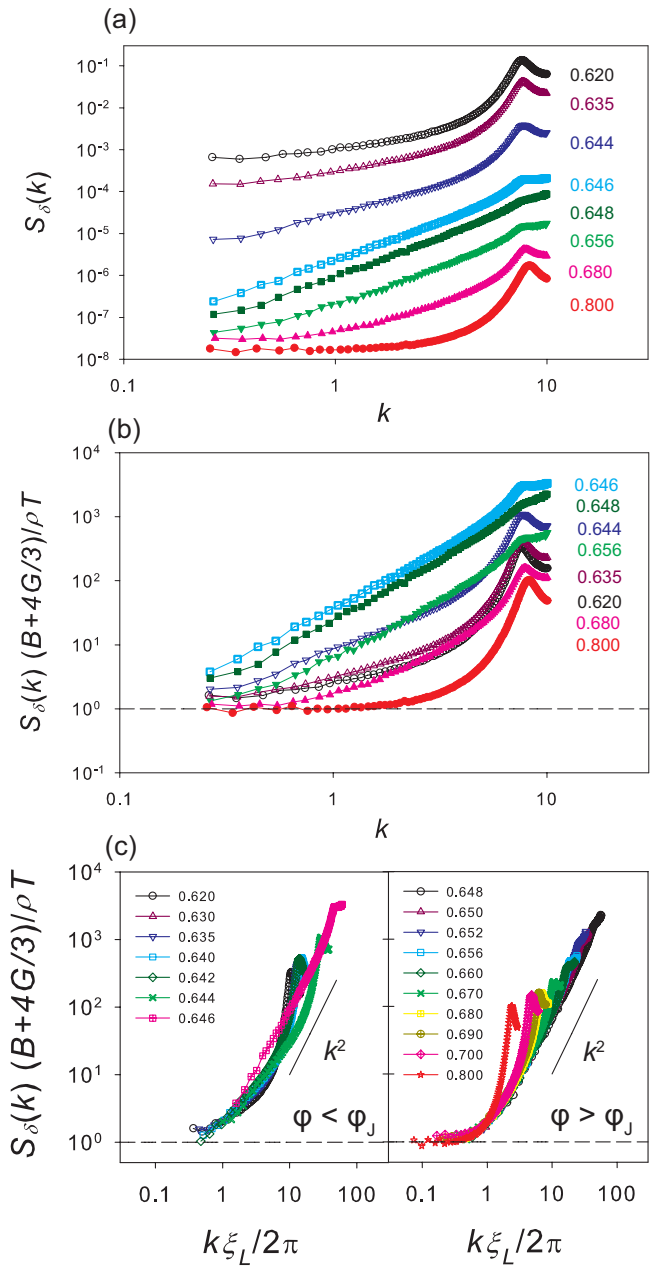,width=8.5cm,clip}
\caption{\label{sdelta} (Color online) 
(a) Fluctuation part of the static structure factor $S_{\delta} (k)$ 
at $T=10^{-8}$ for various densities. 
(b) Same data with vertical axis scaled with 
the factor $(B+ \frac{4}{3} G)/(\rho T)$ expected from 
the usual plane-wave behavior. 
(c) The horizontal axis is also rescaled 
to obtain the best data collapse and extract 
the longitudinal length scale $\xi_L$, Eq.~(\ref{fscale}).}
\end{figure}

We first discuss the behavior of $S_{\delta}(k)$ in connection with 
the bulk modulus. In practice, we
have calculated $S_{\delta}(k)$ in the following way. 
We first calculate the density field $\rho_{\vec k}$ for each instantaneous 
configurations of particles, 
at the lattice points of the first Brillouin zone, 
${\vec k} = (k_x, k_y, k_z)$ in $k$-space 
where $k_{\alpha} = n_{\alpha} \pi/L$ for  $\alpha = x, y, z$ and 
$n_{\alpha}$ an integer. We then perform a time average to obtain 
$\ave{\rho_{\vec k}}$ for each Fourier component. 
Using $\rho_{\vec k}$ and $\ave{\rho_{\vec k}}$ we calculate 
$\delta \rho_{\vec k}$ and obtain $S_{\delta}({\vec k})$ 
in a straightforward manner. We finally perform a circular 
average to obtain the desired $S_{\delta}(k)$.  

We plot the numerical results for 
$S_{\delta}(k)$ at $T=10^{-8}$ and various densities 
in Fig.~\ref{sdelta}. 
The overall amplitude of $S_{\delta}(k)$ strongly decreases 
when the system is compressed. This is expected, because 
particles move less and less when density is increased \cite{ikeda2},
and the overall amplitude of the fluctuations gets smaller.  
At all densities, we also find that $S_\delta(k)$ has a well-defined 
$k \to 0$ limit, and that it increases strongly with $k$, with a well-defined 
first diffraction peak corresponding to the interparticle 
distance at $k \approx k_0 \equiv 2 \pi / \sigma$, 
which reflects the liquid-like structure of amorphous packings at 
the particle scale. 

Two useful informations are contained 
in these structure factors. We first concentrate 
on the $k \to 0$ limits, where the relation Eq.~(\ref{fdtsk}) 
derived from continuum theory is expected to become valid,  
\begin{eqnarray}
\lim_{k \to 0} S_{\delta}(k) = \frac{\rho T}{B+\frac{4}{3}G}. \label{fdtskl}
\end{eqnarray}
To verify this fluctuation formula, we use this expression and 
replot the same data 
in a scaled form, observing the $k$-dependence of the quantity
$S_{\delta}(k) (B + \frac{4}{3}G)/(\rho T)$, as shown in 
Fig.~\ref{sdelta} (middle). 
The bulk and shear moduli are obtained from independent measurements, 
shown in Fig.~\ref{bulk}. Note that since our systems are characterized 
by large $B/G$ ratio, as discussed in Sec.~\ref{mechanical}, 
it means that the bulk modulus gives the major 
contribution to the $(B + \frac{4}{3} G)$ factors in all 
these expresssions, and the term $G$ is almost always negligible. 

The numerical results show that the quantity  
$S_{\delta}(k) (B + \frac{4}{3}G)/(\rho T)$ clearly approaches 
unity as $k \to 0$, as it should. This observation implies that  
standard equilibrium relations between the mechanical moduli and 
static structure factors are valid at large lengthscales for jammed
packings, provided the appropriate fluctuating part of 
the density fields are analysed, instead of the total $S(k)$.  

The rescaled plot shows moreover that $S_\delta(k)$ not only contains 
useful information at $k \to 0$, but that its finite $k$ behaviour is also 
relevant. The conventional elasticity expression in Eq.~(\ref{fdtsk})
does not provide any useful $k$-dependence. 
In other words, when all the longitudinal waves are the plane waves predicted 
by conventional elasticity theory, the quantity 
$S_{\delta}(k) (B + \frac{4}{3}G)/(\rho T)$ should be unity.
Thus, deviations from unity can be interpreted as an indicator 
for the breakdown of the description by the usual plane wave with the above
dispersion relation. Interestingly
we find that the rescaled data 
in Fig.~\ref{sdelta} (middle) show finite-$k$ deviations 
that depend strongly on the density, and are maximal at the 
jamming density, so that the deviations from conventional
elasticity have a remarkable non-monotonic density 
dependence. Interestingly, near the jamming density, 
$\phi \approx \phi_J$, we observe a very clear power law behaviour, 
$S_{\delta}(k) \propto k^2$. 

To characterize quantitatively these deviations and their apparent
relation with the jamming criticality, we propose the
following scaling analysis of these data. The above description 
suggests the existence of a non-trivial correlation 
lengthscale, $\xi_L$, separating two distinct behaviours: 
 $S_{\delta}(k \xi_L \ll 1) (B + \frac{4}{3}G)/(
\rho T) \approx 1$, and $S_{\delta}(1 \ll k \xi_L \ll k_0 \xi_L) 
(B + \frac{4}{3}G)/( \rho T) \propto k^2$. Therefore, 
we determined numerically the longitudinal correlation
lengthscale $\xi_L$ assuming the scaling form
\be
S_\delta(k) \approx \frac{\rho T}{B + \frac{4}{3} G} F(k \xi_L),
\label{fscale}
\ee
where $F(x)$ is a scaling function of the form 
$F(x \ll 1) = 1$ and $F(x \gg 1) \propto x^2$.
Physically, this expression implies that 
$\xi_L$ is a characteristic length scale below which the 
conventional elasticity description of longitudinal particle displacements
breaks down. A diverging correlation length $\xi_L$ implies 
that the usual plane wave description does not apply at any length scale.  
   
The results of this scaling analysis are shown in 
Fig.~\ref{sdelta} (bottom). The data collapse is acceptable, 
but it is difficult to confirm its validity, because the 
obtained lengthscale $\xi_L$ is quite large, and 
a larger system size would be required for a more accurate 
determination of this quantity, especially close to $\phi_J$
at very low temperatures. The evolution of the obtained 
longitudinal lengthscale $\xi_L$ with $\phi$ and $T$ is discussed below in 
Sec.~\ref{lengthscales}. 

\subsection{Transverse fluctuations}
\label{transverse}

\begin{figure}
\psfig{file=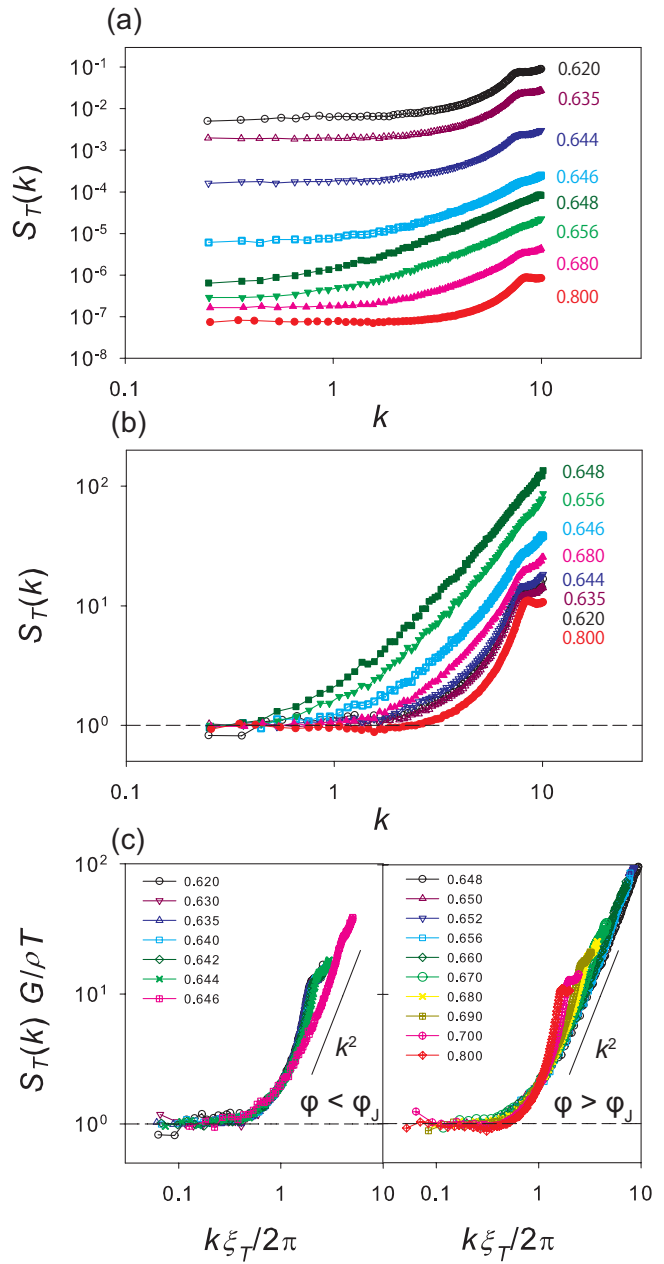,width=8.5cm,clip}
\caption{\label{sdelta2} (Color online) 
(a) Transverse part of the displacement 
structure factor $S_{T} (k)$ at $T=10^{-8}$ for various densities. 
(b) Same data with vertical axis scaled with 
the factor $G/(\rho T)$ expected from the usual plane-wave behavior. 
(c) The horizontal axis is also rescaled 
to obtain the best data collapse and extract 
the transverse length scale $\xi_T$, Eq.~(\ref{hscale}).}
\end{figure}

We now repeat the analysis of Sec.~\ref{longitudinal} for the 
evolution of $S_T(k)$ in connection with the shear modulus. 
We plot $S_T(k)$ at $T=10^{-8}$ and various densities in 
Fig.~\ref{sdelta2} (top). 
As for $S_\delta(k)$, we find that the 
overall amplitude of $S_T(k)$ decreases rapidly with the density,
with an overall wavevector dependence similar to the one 
of $S_\delta(k)$.  

We now use the fluctuation formula for the transverse 
fluctuations~\cite{klix} 
\begin{eqnarray}
\lim_{k \to 0} S_T(k) = \frac{\rho T}{G},
\end{eqnarray}
which is the method we have employed to measure the shear modulus
presented in Fig.~\ref{bulk}. Therefore, by construction, when we replot 
the quantity $S_{T}(k) (G/\rho T)$ in Fig.~\ref{sdelta} (middle),
the data extrapolate to unity when $k \to 0$. 
 
This figure shows again that the deviations from conventional 
elasticity have a striking non-monotonic density dependence,
and that deviations are maximal when $\phi$ is close to $\phi_J$. 
Following the analysis of longitudinal fluctuations, we again
hypothesize a scaling behavior for $S_T(k)$: 
\be
S_T (k) \approx \frac{\rho T}{G} H(k \xi_T),
\label{hscale}
\ee
where $H(x)$ is a scaling function of the form 
$H(x \ll 1) = 1$ and $H(x \gg 1) \propto x^2$.
We show in Fig.~\ref{sdelta2} (bottom) a collapse of the numerical 
data according to Eq.~(\ref{hscale}) which allows us to 
determine numerically a lengthscale $\xi_T$
which delimits the validity of the usual plane wave description 
of transverse fluctuations of the particle displacements.
This critical scaling law implies again that the usual plane wave description 
does not apply at any length scale when $\xi_T$ diverges, 
which is expected exactly at the jamming transition. 

\subsection{Transverse and longitudinal lengthscales}
\label{lengthscales}

\begin{figure}
\psfig{file=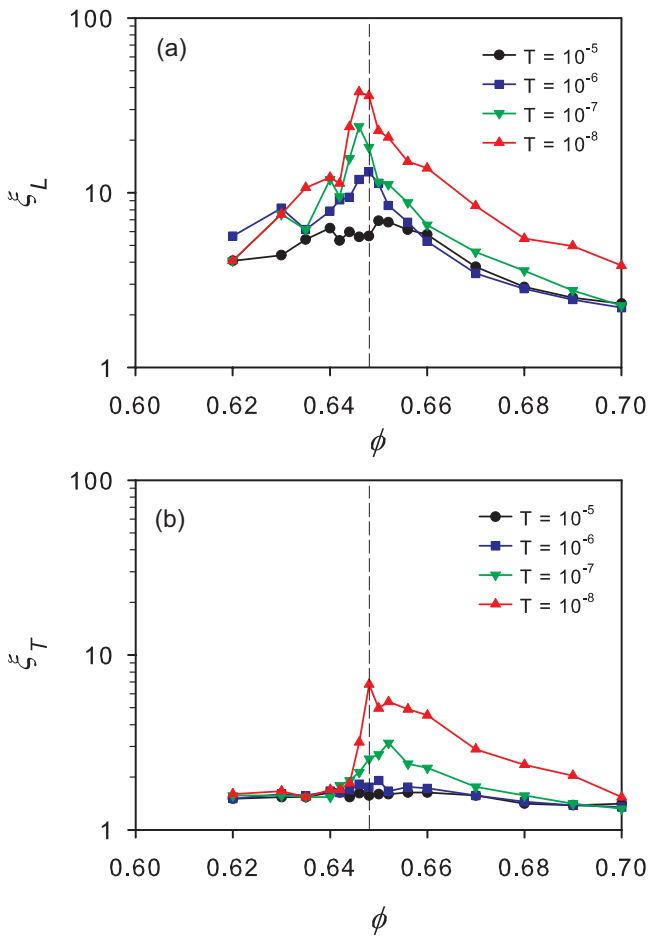,width=8.5cm,clip}
\caption{\label{xi_long} (Color online) 
Density dependence of $\xi_L$ (a) and $\xi_T$ (b)
for various temperatures. The near-critical 
non-monotonic density dependence observed for $T=10^{-8}$ 
becomes a smooth variation for $T=10^{-5}$ when jamming 
criticality is smeared by thermal fluctuations. The usual 
plane wave description 
of vibrational motion is therefore excellent away from criticality, 
$T \gtrsim 10^{-6}$ and/or $|\phi - \phi_J| \gtrsim 0.02$.} 
\end{figure}

We have performed the scaling analysis outlined in Secs.~\ref{longitudinal} 
and \ref{transverse} for different temperatures and have been able 
to extract the density and temperature 
dependences of the characteristic length scales 
$\xi_L$ and $\xi_T$. The results are presented in Fig.~\ref{xi_long}.

Both lengthscales behave qualitatively similarly. 
At $T=10^{-8}$, they strongly depend on the density:
they have a clear maximum for $\phi \approx \phi_J$ 
and become of order unity far away from the jamming transition.  
In particular, we observe that $\xi_L$ becomes 
comparable to the system size ($L=36.2$) for $T=10^{-8}$, which 
explains why its numerical determination is difficult.
By contrast, the maximum value reached by $\xi_T$ is more modest
(or order $\xi_T \approx 10$), explaining why the 
data collapse for $S_T(k)$ is more convincing than the one for $S_\delta(k)$. 
 
When the temperature is increased, the density maximum observed 
for the correlation lengthscales
is much less pronounced, and nearly disappears when $T=10^{-5}$
where $\xi_L$ and $\xi_T$ have uninteresting density dependences
and remain microscopic, $\xi_{L,T} \approx 2-5$. 
The conclusion is that in this regime, 
which is relevant for a number of experimental situations
for colloidal systems, continuum mechanics actually 
represents an excellent approximation down to microscopic length scales. 
In other words, ``anomalous'' or ``soft'' modes, which 
exist over arbitrary length and frequency scales at the jamming 
transition where correlation lengthscales and timescales 
are infinite, are strongly suppressed by moving away from jamming criticality.  

We now compare our measurements of the lengthscales 
$\xi_L$ and $\xi_T$ to similar lengthscales measured earlier 
in the literature. A first relevant comparison is with the 
results in Refs.~\cite{silbert1,silbert2} where a characteristic 
length scale for longitudinal and transverse plane waves at 
a specific frequency $\omega^\ast = \omega^{\ast}(\phi)$ for $\phi > \phi_J$ 
and $T=0$ were measured. In this approach,
$\omega^\ast$ is a characteristic frequency 
where the vibrational density of state exhibits anomalous (nearly 
frequency-independent) behavior~\cite{epitome,wyart}. 
The obtained longitudinal and transverse length scales $\xi_L^{\ast}$ 
and $\xi_T^{\ast}$ measured from this protocol 
are predicted to diverge as $\xi_L^{\ast} \propto (\varphi - \varphi_J)^{-0.5}$ 
and $\xi_T^{\ast} \propto (\varphi - \varphi_J)^{-0.25}$~\cite{silbert1}, 
and the latter behavior is directly confirmed by numerical 
measurements (data for $\xi_L^\ast$ were not shown). 

Although it is tempting to assume that $\xi_L$ and $\xi_T$ 
have similar physical content as $\xi_L^{\ast}$ and 
$\xi_T^{\ast}$, power law fits to our measurements 
yield exponents that are not consistent with the predicted 
$0.5$ and $0.25$ (we find that 0.7 and 0.5 fit our data better). 
However, the lengthscales observed in our measurement is so large 
that their precise determination is challenging. 
Much larger systems are needed to fully settle this issue. 
Furthermore, our lack of knowledge about the precise form 
of the scaling functions $F(x)$ and $H(x)$ may 
also affect the quality of our estimates for these lengthscales. 
We wish to raise the possibity that the two sets of 
lengthscales are not fully equivalent because we directly defined 
characteristic lengthscales where the usual 
plane wave descriptions stop working, 
whereas Silbert {\it et al.} focused on a specific, density-dependent frequency 
$\omega^{\ast}$ \cite{silbert1,silbert2}.

In a more recent work, Wang {\it et al.}~\cite{xu} 
have determined numerically the so-called Ioffe-Regel frequencies and 
lengthscales. These are respectively  
defined as the timescales and lengthscales characterizing the 
disappearance of plane waves in the collective 
dynamic structure factors. We may expect that the lengthscales 
$\xi_L$ and $\xi_T$ that we have defined above carry a similar physical
content to the Ioffe-Regel lengthscales $l_L$ and $l_T$ analysed 
in Ref.~\cite{xu}. Although some of the numerical 
results of Wang {\it et al.} are consistent with ours, their final conclusions 
differ qualitatively from ours. In particular, their analysis
suggests that in the low-temperature limit $l_T$ diverges 
at $\varphi = \varphi_J$ and transverse 
plane waves do not exist in the hard sphere regime 
$\varphi < \varphi_J$ and $T \to 0$. Instead, we observe that $\xi_T$ becomes 
finite on both sides of the jamming transition. We suspect that 
their interpretation is incorrect and stems from extrapolating 
numerical measurements performed at $\phi > \phi_J$ to the 
hard sphere glass. Our results show instead that longitudinal 
and transverse vibrations do exist in the hard sphere regime, 
and the associated lengthscales and timescales actually become  
microscopic as the density is decreased away from $\phi_J$.
Contrary to the claim in Ref.~\cite{xu}, the hard sphere
glass is not qualitatively different from other amorphous materials.

\section{Hyperuniformity of the configurational 
structure factor}
\label{hyperuniformity}

In this section, we analyse the structure factor associated to
averaged particle positions and show that this configurational part 
$S_0(k)$ reveals a nearly-hyperuniform 
behaviour at large length scales. We also show how the 
analysis of $S_0(k)$ allows us to elucidate the physical content
of the related nonequilibrium index and nonequilibrium lengthscale 
measured in amorphous materials. 

\subsection{Hyperuniformity is unrelated to jamming 
criticality}

In Sec.~\ref{definitions}, we decomposed the structure factor 
$S(k)$ into a fluctuating part, analysed in Sec.~\ref{longitudinal},
and a configurational part $S_0(k)$, which is the subject of the 
present section.  

\begin{figure}
\psfig{file=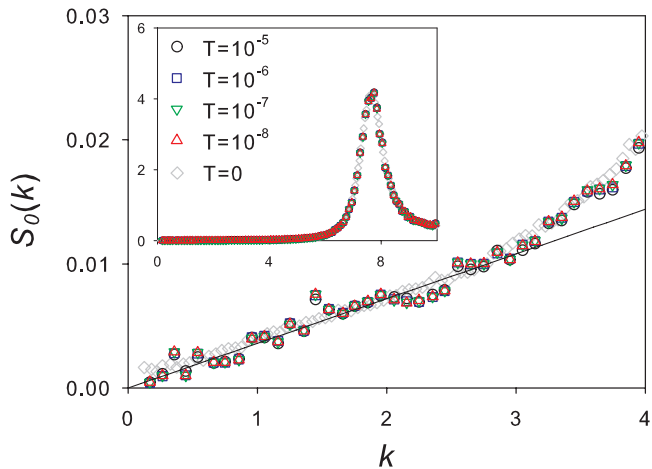,width=8.5cm,clip}
\caption{\label{s0temp} (Color online) 
Configurational part of the static structure factor $S_0(k)$ at 
$\varphi=0.652 > \phi_J$ and several temperatures.  
Circles, squares, and upper and lower triangles are 
represent the data obtained at finite 
temperatures for $N=64,000$, diamonds indicate data
obtained at $T=0$ for $N=512,000$. 
Solid line indicates the hyperuniform behaviour  
of density fluctuations at large scale, $S_0(k) \propto k$. 
The hyperuniformity appears essentially independent of temperature.
The inset is the zoom out of the main panel.}   
\end{figure}

The first question we ask is whether the configurational part 
$S_0(k)$ is sensitive to the temperature.
In Fig.~\ref{s0temp}, we show numerical results for $S_0(k)$ 
at a constant density $\varphi = 0.652$ (slightly above
$\phi_J$) at several temperatures from $T=10^{-8}$ up to $T=10^{-5}$. 
The inset shows that $S_0(k)$ has the usual wavevector dependence 
of a liquid structure factor, with a broad first diffraction peak
near $k \approx k_0$. Clearly, the temperature dependence appears  
negligible in this representation.

In the main panel, we zoom on the low-wavevector behaviour in order to 
reveal a possible effect of the thermal fluctuations. 
Within the
accuracy of these computations, we find again no visible 
temperature dependence for $S_0(k)$. This clearly confirms
that thermal fluctuations mainly contribute to the fluctuating part 
of the density fluctuations, whereas the
averaged component of density fluctuations is essentially unaffected
by the temperature, at least for the range of wavevectors shown 
in Fig.~\ref{s0temp}. 
We expect more changes to occur at very large 
wavevectors, $k \gg k_0$, where the sharp features associated to the
pair correlation function at contact produce long-ranged 
oscillations \cite{ikeda4}. 

An interesting behaviour is observed for the low-$k$ behaviour of 
$S_0(k)$ in Fig.~\ref{s0temp}. In the linear scale chosen for
representating these numerical measurements, we observe 
that $S_0(k) \propto k$, for $k \lesssim 2$. 
This ``anomalous'' linear behavior with an apparent vanishing of $S_0(k\to0)$ 
has been termed hyperuniformity \cite{donev2,torquato2003}. 
Hyperuniform density fluctuations have been reported in simulations 
of sphere packings at the jamming transition both 
numerically~\cite{donev2,zac,berthier,silbert3} and 
in experimentally-constructed granular packings~\cite{berthier}.
In colloidal systems, signs of hyperuniformity 
are much weaker~\cite{weeks1,kurita,dreyfus}.

The important conclusion that we can draw from the 
absence of temperature dependence in the data shown in 
Fig.~\ref{s0temp} is that hyperuniformity appears extremely robust 
against thermal fluctuations, and can in fact easily be observed 
even for our highest studied temperature $T=10^{-5}$. 
This observation is in striking contrast with all other 
observations reported earlier in this paper related to
the jamming criticality. Whereas quantities such as 
mechanical moduli and correlation lengthscales are rapidly smeared
out by thermal fluctuations, hyperuniformity appears rather 
insensitive to temperature. This strongly suggests 
that {\it jamming criticality and hyperuniformity are unrelated concepts}
and have distinct physical origins.

We emphasize that the decomposition of 
the structure factor $S(k)$ as the sum of two terms $S_0(k)$ and 
$S_\delta(k)$ indicates that the total structure factor is related to 
the isothermal compressibility only through $S_\delta(k)$ whose 
wavevector dependence shows no anomalous dependence, see Fig.~\ref{sdelta}. 
On the other hand, we find that $S_0(k)$ is characterized
by a hyperuniform linear behaviour at low wavevector, but this 
configurational contribution is unrelated to the compressibility. 
Thus, we conclude that hyperuniformity (related to $S_0$) 
cannot be a logical consequence of a vanishing 
compressibility (related to $S_\delta$) of the packing. Of course, 
to observe hyperuniformity in the total structure factor $S(k)$ 
rather than in the configurational part $S_0(k)$, 
it is necessary that $S_\delta(k) \ll S_0(k)$, which happens when the 
compressibility becomes very small. Working close to the 
jamming transition is therefore {\it a practical rather than 
a fundamental issue}, which is no longer needed when 
working directly with $S_0(k)$. In that sense, the discovery that 
some amorphous materials become hyperuniform very close to 
the jamming transition may appear coincidental \cite{donev2}.  

These findings are experimentally relevant, because they mean 
that hyperuniformity, unlike jamming criticality, can actually 
be observed across a large region of the $(T,\phi)$ phase diagram 
for soft colloids. Quite surprisingly, the experimental literature seems
to suggest exactly the opposite since a number of experiments 
have reported signatures of jamming criticality in 
soft colloids~\cite{yodh1,vestige}
(which, we argue, are actually taken far from criticality), 
whereas only weak signs
of hyperuniformity have been reported~\cite{weeks1,kurita,dreyfus}
(which, we argue, should be easily observable).
This might be due to the difficult experimental constraint 
that measuring $S_0(k)$ at low $k$ requires detecting the position 
of a large number of particles with a large precision. 

\subsection{Density dependence and deviations from strict
hyperuniformity}

Having established that temperature does not affect 
much the observation of hyperuniformity, we then discuss the 
effect of the density by analysing data obtained directly 
at $T=0$. This approach is useful, because it allows us to 
disentangle temperature and density effects. In addition, 
we can study at $T=0$ much larger systems in order to analyse 
whether hyperuniform
behaviour can be observed over arbitrarily large lengthscales. 
To this end, we employ a distinct measurement technique 
of $S_0(k)$ at $T=0$ with a larger number of particles, $N=512,000$, 
and use 8 independent packings to reduce the statistical error. 
Our numerical methodology was described in Sec.~\ref{model}. 
 
Using this larger systems at $T=0$, we confirm in 
Fig.~\ref{s0temp} that $S_0(k)$ for this density is in excellent 
agreement with the finite temperature results obtained with the smaller
packings, although of course the statistical error is greatly 
reduced. The agreement between these two independent sets of data confirms
that $S_0(k)$ is largely independent of temperature.
 
\begin{figure}
\psfig{file=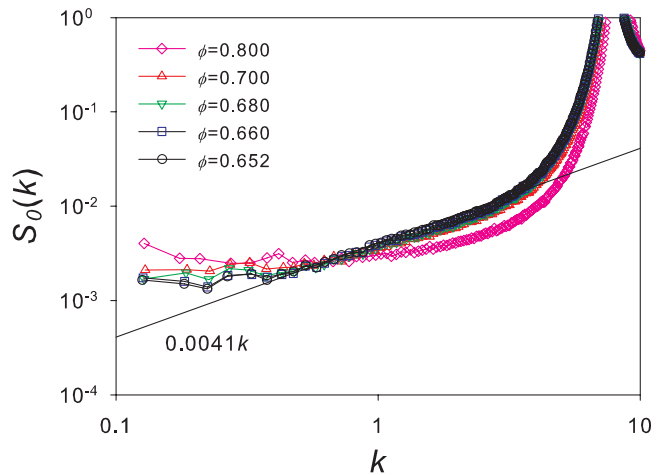,width=8.5cm,clip}
\caption{\label{s0dens} (Color online) Evolution of $S_0(k)$ with the density 
above the jamming transition in a log-log representation.
These data are obtained by averaging over
8 independent packings with $N=512,000$ particles. A clear hyperuniform 
linear $k$-dependence (shown as a full line) 
is obtained over a broad range of wavevectors when 
$\phi \lesssim 0.7$ with only weak density dependence, but the data
saturate to a finite value as $k \to 0$.}
\end{figure}

Having shown that temperature plays no role, we can now
analyse the density dependence of these observations.
In Fig.~\ref{s0dens} we plot $S_0(k)$ measured at $T=0$
at various densities between $\phi=0.80$ much above jamming, 
down to $\phi=0.652$ just above $\phi_J$. We now use a log-log 
representation of the results. 
At $\varphi=0.80$, $S_0(k)$ is almost constant 
over a large range of wavevectors, $S_0(k) \approx 4 
\cdot 10^{-3}$ for $k \lesssim 1$, and a hyperuniform behaviour cannot 
be observed.  This behaviour of over-compressed packings is 
consistent with observations made in other glass-forming materials,
such as simple Lennard-Jones glasses,  
where hyperuniformity is not observed either~\cite{silbert3}. 

For $\varphi=0.7$, a linear behavior, $S_0(k) \approx 0.004 k$, already 
appears in wide $k$ region, even though $|\phi - \phi_J| \approx 0.055$ 
is still quite large. More surprising is the observation that 
the data for $\phi = 0.652$, 0.66 and 0.68 
(respectively corresponding to $|\phi-\phi_J| \approx 0.0063$, 0.015, and 0.034)
are essentially the same, 
and are characterized by a broad range of wavevectors with 
linear dependence, $S_0(k) \approx 0.004 k$, although the data saturate
at very low $k$ to a finite value $S_0(k\to0) \approx 1.4 \cdot 10^{-3}$. 
These results indicate that hyperuniformity is a robust 
feature of $S_0(k)$, in the sense that it is weakly dependent on the density
and does not require fine-tuning the volume fraction 
to the jamming density $\phi_J$, confirming that the two concepts are distinct. 
  
However, it should also be noted that a strict hyperuniformity 
$S(k) \propto k$ can not be observed down to arbitrarily small wavevectors, 
and deviations appear below $k \approx 0.4$, which corresponds to 
a large lengthscale $\approx 15 \sigma$.   
This surprising saturation effect has not been 
reported before, although we notice that previous 
literature~\cite{donev2,silbert3} 
indicates that the smallest $S_0(k\to 0)$ values achieved 
in computer simulations are always of the order $10^{-3}$ or more, 
which is consistent with our own results.  This saturation 
would not be observed if the data in Fig.~\ref{s0dens} where plotted 
in a linear scale. 

Our analysis shows that this saturation effect is clearly not due 
to thermal fluctuations (we work at $T=0$). This does not stem
from sample-to-sample fluctuations either, because all 8 samples
show a saturation of similar amplitude. Finally, 
the saturation does not seem to depend on density, 
at least for $\phi > \phi_J$.  
We cannot access the regime $\phi < \phi_J$ using energy minimization, but
we note that a marked density dependence was reported 
for the structure factor of hard spheres approaching 
the jamming density from below in Ref.~\cite{donev2}. However, 
the total structure factor $S(k)$ was measured in that study, 
which contains a density-dependent contribution associated to 
the fluctuating part $S_\delta(k)$, which vanishes as $\phi_J$ is approached. 
It would therefore be very interesting 
to measure directly $S_0(k)$ in the hard sphere glass 
for very large system sizes. We have performed exploratory simulations 
with moderate system sizes in this regime and find a weak density dependence 
of $S_0(k)$ when $\phi > 0.62$, but larger systems are needed to 
analyse more finely the behaviour at very small $k$.  

Finally, we remark that it is difficult to 
provide a physical explanation for the existence of the 
observed deviations from  
strict hyperuniformity, mainly because there is no deep physical
reason to expect perfect hyperuniformity in these 
systems in the first place. 
Among possible factors that could be investigated are the role 
of a finite density of rattlers in the packings and the role 
of the specific protocol that is used to prepare the particle 
packings, which both could influence the measured value of $S_0(k \to 0)$. 
Such studies are beyond the scope of the present work. 

\subsection{Analysis of the ``nonequilibrium index''}

We showed in Sec.~\ref{longitudinal}
that the isothermal compressibility is directly 
related, for {\it solids} at thermal equilibrium, to the fluctuating
part of the structure factor via Eq.~(\ref{fdtsk}). A direct consequence
is that the compressibility is thus not related to the total
structure factor $S(k)$ via the relation valid for equilibrium {\it fluids}, 
Eq.~(\ref{liqb}). To quantify the difference between fluid and solid
states, the concept of a ``nonequilibrium index'', $X$, was 
introduced and studied both for hard sphere
glasses \cite{torquato2} and for other types of amorphous 
materials \cite{torquato3}. 

We now show that the decomposition provided above for the structure
factor allows us to elucidate the physical content of $X$. 
Using the notations introduced in the present work, 
the nonequilibrium index $X$ is defined as~\cite{torquato2} 
\begin{eqnarray}  
X \equiv \lim_{k \to 0} \frac{S(k) B}{\rho T} - 1 \label{neqx}
\end{eqnarray} 
By construction, $X=0$ for a fluid at thermal equibrium, see 
Eq.~(\ref{liqb}). 
Since $X$ is defined as the ratio between fluctuations and 
response functions, its functional form is also reminiscent 
of the effective temperature and fluctuation-dissipation ratio 
that characterize the nonequilibrium of aging and driven 
glasses \cite{leticia}. The main difference between the two types of 
quantities is that $X$ refers to static fluctuations, whereas
effective temperatures are defined from time-dependent 
correlation and response functions.  
The non-zero value measured for $X$ in 
hard sphere glasses was interpreted as a demonstration that the 
``{\it jammed glassy state is fundamentally nonequilibrium in 
nature}''~\cite{torquato2}. The simulations indicate that 
$X$ grows rapidly when hard sphere glasses are compressed towards 
$\phi_J$, or when the temperature is decreased below the glass transition 
temperature $T_g$ of model glass-forming systems, 
such as Lennard-Jones and Dzugutov glasses~\cite{torquato3}. 

\begin{figure}
\psfig{file=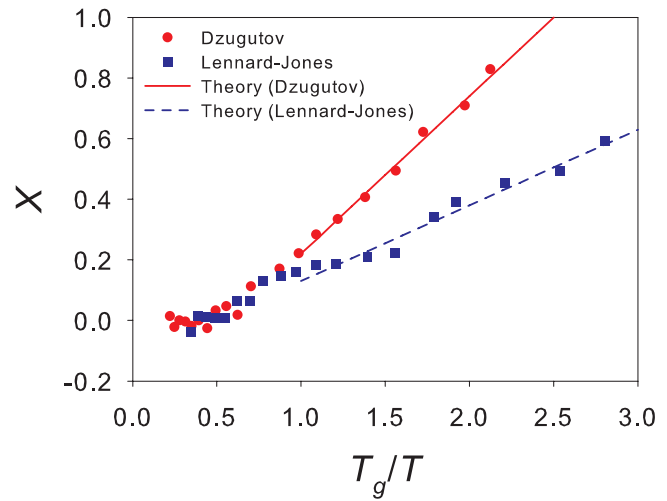,width=8.5cm,clip}
\caption{\label{noneqX} (Color online) Temperature 
dependence of the nonequilibrium index $X$ measured 
in two model glasses (symbols). Full lines 
are from Eq.~(\ref{Xglass}), the prediction obtained by 
assuming that equilibrium fluctuation relations are 
satisfied for the glass.
}
\end{figure}

The decomposition of the structure factor proposed in 
Eq.~(\ref{sum}) provides us with two important informations. 
First, we have shown that equilibrium fluctuation relations 
are perfectly obeyed in the solid phase for static quantities. 
This result implies that the nonequilibrium nature of the glass cannot 
be revealed by a fluctuation formula based on static density
fluctuations and suggests, in fact, that the introduction 
of a ``nonequilibrium'' index to characterize static density 
fluctuations is unnecessary. This conclusion is in qualitative
agreement with the two-temperature scenario for the nonequilibrium 
dynamics of glasses, where short-time fluctuations and response
are typically found to obey equilibrium 
fluctuation-dissipation relations~\cite{leticia,jorge}.  

Second, the combination of Eqs.~(\ref{sum}, \ref{fdtsk}) provides 
predictions for the leading behaviour of the nonequilibrium 
index in various systems. 
For glass-forming models with continuous interactions, we can assume 
that $S_0(k \to 0)$ and the bulk modulus are weakly temperature 
dependent deep in the glass phase~\cite{ikeda4}, 
so that in the low-temperature limit, one gets 
\be 
X(T \ll T_g) \approx \frac{S_0(k\to0) B}{\rho T} \propto \frac{1}{T}.
\label{Xglass}
\ee 
In Fig.~\ref{noneqX}, we confirm that the low-temperature behaviour
of the nonequilibrium index measured numerically in Ref.~\cite{torquato3}
is consistent with our prediction 
in Eq.~(\ref{Xglass}) that it should diverge as $1/T$ as $T \to 0$. 
This leading temperature behaviour stems from the fact that 
$S(k)$ in the definition of $X$ in Eq.~(\ref{neqx}) contains 
a `frozen' contribution, $S_0$, which does not vanish at $T=0$. 

For hard sphere glasses, the leading asymptotic behaviour 
of the nonequilibrium index depends strongly of the hypothesis 
made regarding the behaviour of $S_0(k)$ very close to $\phi_J$. 
Assuming that hyperuniformity is only approximate, as we report
in Fig.~\ref{s0dens}, 
one would then predict that $X$ in Eq.~(\ref{neqx}) 
is dominated by the divergence of the bulk modulus, yielding 
$X \approx (\varphi_J - \varphi)^{-2}$. In Ref.~\cite{torquato2}
a linear decrease, $S_0(k \to 0) \sim (\phi - \phi_J)$, was assumed 
for $S_0$, which turns into a different divergent behaviour,
$X \sim (\varphi_J - \varphi)^{-1}$, for the nonequilibrium index. 
Numerically, we expect that $X$ exhibits a crossover 
between these two power law regimes as $\phi_J$ is approached, 
which could be difficult to analyse. 

\subsection{Analysis of the ``nonequilibrium lengthscale''}

\begin{figure}
\psfig{file=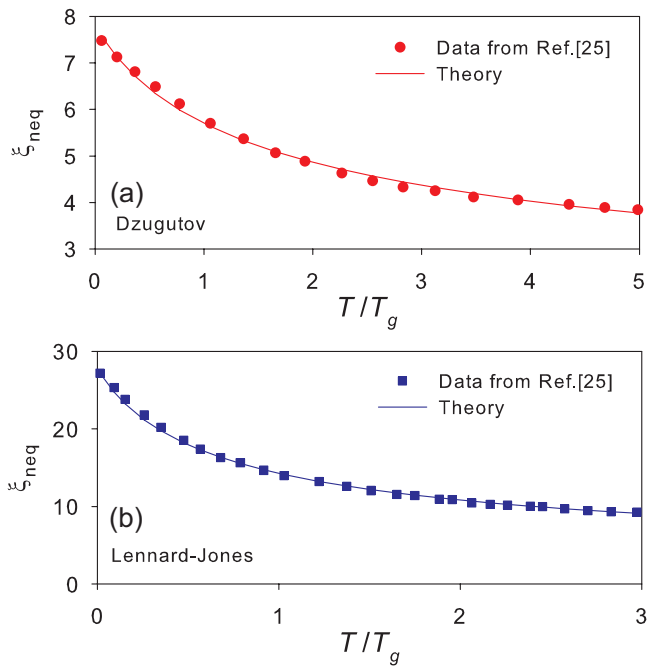,width=8.5cm,clip}
\caption{\label{neql} (Color online) 
Temperature dependence of the nonequilibrium 
lengthscale $\xi_{\rm neq}$ measured in two model glasses
(symbols). Full lines are from 
Eq.~(\ref{neqxi_ours}), the prediction obtained 
by assuming the equilibrium fluctuation relations 
are satisfied for the glass. 
}
\end{figure}

We finally discuss the concept of a nonequilibrium lengthscale,
$\xi_{\rm neq}$, defined again for amorphous materials 
from the behaviour of the static structure structure. 
The nonequilibrium lengthscale is defined as~\cite{torquato3} 
\begin{eqnarray}  
\xi_{\rm neq} \equiv [-c(k \to 0)]^{1/d} \approx 
[\rho S(k \to 0)]^{-1/d}, \label{neqxi}
\end{eqnarray} 
where $c(k)$ is the direct correlation function~\cite{hansen}
and $d$ is the dimensionality of the system. In the final part of 
Eq.~(\ref{neqxi}), we have assumed that $S(k\to0) \ll 1$. 
In computer simulations, it is found that $\xi_{\rm neq}$ grows as 
temperature is decreased below $T_g$ in model glasses \cite{torquato3}
and saturates to a finite value as $T \to 0$, 
whereas it is predicted to diverge as $\phi \to \phi_J$ in hard sphere 
glasses \cite{torquato2}. As such, it is interpreted as 
a growing static length scale that is potentially 
relevant to characterize the structure of the glassy state.
In this view, hard sphere glasses would therefore be somewhat ``special'' since 
they would have a diverging static lengthscale, whereas glasses with 
continuous interactions would exhibit a non-diverging static lengthscale. 

Because this lengthscale directly follows from the low-$k$ behaviour 
of the structure factor, our decomposition (\ref{sum}) into two 
distinct contributions is again relevant to understand the 
physical content of the nonequilibrium lengthscale, which we can rewrite 
\be
\xi_{\rm neq}  = \left[ 
\rho S_0(k\to0) + \frac{\rho^2}{B+\frac{4}{3}G} T \right]^{-1/d}.
\label{xiT}
\ee  
For Lennard-Jones and Dzugutov models, we again 
expect that $S_0(k\to 0)$ and the mechanical moduli are weakly dependent
on temperature far below $T_g$, so that the leading temperature dependence of
the nonequilibrium lengthscale is transparent in 
Eq.~(\ref{xiT}), and should be of the form
\begin{eqnarray}  
\xi_{\rm neq} \approx ( a + b T )^{-1/d}, \label{neqxi_ours}
\end{eqnarray} 
where $a$ and $b$ are some constants. In Fig.~\ref{neql}, we confirm that
this prediction describes the numerical data very well
for two glass-formers, showing that the growth of the 
nonequilibrium lengthscale with decreasing temperature 
can in fact be fully understood by assuming that 
density fluctuations obey equilibrium fluctuation formula. 
In essence, therefore, the growth of the nonequilibrium lengthscale
in the glass phase reflects the competition between the configurational 
and fluctuating parts of the static structure factor, which have different
temperature dependences: the former is essentially constant and reflects
the `inherent' structure of the glass, the latter stemming from 
vibrational motion and is thus proportional to temperature, 
as captured in Eq.~(\ref{neqxi_ours}).

For hard sphere glasses, the behaviour of the 
nonequilibrium length would again depend sensitively 
on the behaviour of $S_0(k)$ near $\phi_J$. Assuming that 
hyperuniformity is only approximate, the nonequilibrium lengthscale would 
grow strongly as the glass phase is entered and the compressibility 
decreases, but its growth would saturate to a value 
$\xi_{\rm neq} \approx [\rho S_0(k\to 0)]^{-1/d} \approx 8.2$, 
using numerical values from Fig.~\ref{s0dens}. 
Interestingly, this saturation value is close to the value 
$\xi_{\rm neq} (T\to0) \approx 7.5$ found 
for the three-dimensional Dzugutov glass-former in Fig.~\ref{neql},
which could support the idea that hard sphere glasses are not 
a ``special'' type of glass-former.
For a strictly hyperuniform hard sphere system, on the other hand, 
the nonequilibrium lengthscale
would diverge as $\phi \to \phi_J$, as predicted in 
Ref.~\cite{torquato2}.  

\section{Conclusion}
\label{summary}

In this work, we have analyzed the density and temperature dependences of 
mechanical moduli and several types of structure factors in a
model system of soft harmonic spheres in the vicinity of the jamming 
transition. 

We have shown that thermal fluctuations very quickly erase several
signatures of the criticality associated to jamming, in agreement with
earlier work related to single particle dynamics \cite{ikeda2}. We showed that 
the bulk modulus, the shear modulus, the longitudinal and transverse 
lengthscales rapidly acquire a `normal' behaviour typical
of ordinary solids, whereas the large lengthscales and timescales
associated to the isostatic jamming critical point are only observed
in a narrow region of the $(T,\phi)$ phase diagram. 
We conclude that most colloidal experiments to date have 
hardly been able to probe the jamming criticality, nor have 
the thermal vestiges of the jamming transition that result from
the existence of non-microscopic lengthscales and timescales
been observed. These conclusions suggest that the soft and 
hard sphere glasses that are commonly studied experimentally
essentially behave as ordinary solids where the usual plane wave description
holds down to small length scales, as concluded from a very recent 
experimental study \cite{keim}. Therefore, we hope that 
our results will encourage further experimental investigations
of these issues, for instance using emulsion droplets \cite{emulsion}, or 
core-shell microgel particles~\cite{peter}. 

A second major finding in our study is that density fluctuations 
for jammed colloidal systems follow the laws of equilibrium 
thermodynamics and their study does not reveal the 
nonequilibrium nature of glasses. Our analysis is based 
on a decomposition of density fluctuations in a configurational 
and fluctuating parts. Whereas the fluctuating part is directly related
to mechanical moduli via equilibrium fluctuation formula, 
we found that the configurational part is essentially independent
of both density and temperature in a rather broad range of parameters.
The decomposition into these two components allows us to elucidate
the behaviour reported in earlier numerical studies for 
the nonequilibrium index and nonequilibrium lengthscales 
characterizing amorphous materials.
Based on these observations, we have suggested that hyperuniformity
observed in the configurational structure factor is unrelated to 
the compressibility, and therefore to the jamming criticality. 

These results raise some interesting questions. It has been 
established numerically that the same jamming criticality is observed
for packings with very different preparation protocols~\cite{pinaki}. 
Our observation
that a strict hyperuniformity is not observed in our packings suggests
that the value of $S_0(k \to 0)$ could very well be affected by the 
nonequilibrium protocol used to prepare packings. One could 
for instance hypothesize that a packing prepared with a slower annealing could
be more hyperuniform than one produced via brutal compression. This raises
the appealing possibility that the nonequilibrium lengthscale
$\xi_{\rm neq}$ measured either at $T=0$ (for continuous potentials)
or at infinite pressure (for hard spheres) truly encodes some non-trivial
information about the glassy state \cite{torquato2}. If correct, it would mean
that it is not really the temperature or density dependences
of $\xi_{\rm neq}$ which truly matter, but rather its evolution for different 
preparation histories. Therefore, we believe that it would be 
interesting to understand better the physical content of this quantity in 
various glassy materials prepared using various thermal histories.

\acknowledgments
We thank S. Teitel for useful discussions, H. Mizuno for helpful 
comments, and E. Corwin for sharing some of his packings with us. 
The research leading to these results has received funding
from the European Research Council under the European Union's Seventh
Framework Programme (FP7/2007-2013) / ERC Grant agreement No 306845, 
and from JSPS KAKENHI Grant Number 26887021.

\end{document}